\documentclass[pra,twocolumn]{revtex4-2}
\usepackage{amsmath,amssymb,mathrsfs}
\usepackage{psfrag}
\usepackage{graphicx}
\usepackage{graphics}
\usepackage{epsfig}
\usepackage{bm}
\usepackage{color}
\usepackage{verbatim,color}
\usepackage{physics}
\usepackage[normalem]{ulem}
\usepackage[acronym]{glossaries}
\newacronym{dr}{DR}{dimensional regularization}
\newacronym{ms}{MS}{minimal subtraction}
\newacronym{eft}{EFT}{effective field theory}

\newcommand{\beq}{\begin{equation}}
\newcommand{\eeq}{\end{equation}}
\newcommand{\beqa}{\begin{eqnarray}}
\newcommand{\eeqa}{\end{eqnarray}}
\newcommand{\ba}{\begin{aligned}[b]}
\newcommand{\ea}{\end{aligned}}

\newcommand{\cblue}{\color{black}}
\newcommand{\cred}{\color{black}}

\begin{document}

\title{On-shell approximation for the s-wave scattering theory}
\author{F. Lorenzi$^{1,2}$, A. Bardin$^{1}$, and L. Salasnich$^{1,2,3}$}
\affiliation{$^{1}$Dipartimento di Fisica e Astronomia "Galileo Galilei", 
Universit\`a di Padova, Via Marzolo 8, 35131 Padova, Italy\\
$^{2}$Istituto Nazionale di Fisica Nucleare (INFN), Sezione di Padova, via Marzolo 8, 35131 Padova, Italy\\
$^{3}$Istituto Nazionale di Ottica (INO) del Consiglio Nazionale delle Ricerche (CNR), via Nello Carrara 1, 50019 Sesto Fiorentino, Italy}

\date{\today}

\begin{abstract}
We investigate the scattering theory of two particles in a generic $D$-dimensional space. For the s-wave problem, by adopting an  on-shell approximation for the $T$-matrix equation, we derive analytical formulas which connect the Fourier transform ${\tilde V}(k)$ of the interaction potential to the s-wave phase shift. In this way we obtain explicit expressions of the low-momentum parameters ${\tilde g}_0$ and ${\tilde g}_2$ of ${\tilde V}(k)={\tilde g}_0+{\tilde g}_2k^2 +...$ in terms of the s-wave scattering length $a_s$ and the s-wave effective range $r_s$ for $D=3$, $D=2$, and $D=1$. Our results, which are strongly dependent on the spatial dimension $D$, are a useful benchmark for few-body and many-body calculations. As a specific application, we derive the zero-temperature pressure of a 2D uniform interacting Bose gas with a beyond-mean-field 
correction which includes both scattering length and effective range. 
\end{abstract}

\maketitle

\section{Introduction} 
One of the main features of the physics of ultracold and dilute atomic gases is their universality, i.e. the fact that the interaction potential, and consequently many physical properties, can be accurately described by only one zero-range interaction parameter: the s-wave scattering length \cite{leggett,stringari}. The flexibility of current experimental techniques prompts novel interest in the nonuniversal behavior of quantum gases. One remarkable example is the possibility of using Feshbach resonances for tuning the s-wave scattering length, and eventually obtain an interaction regime in which the next-to-leading order term in the low momentum expansion of the potential, i.e. the effective range $r_s$, becomes relevant \cite{stoof}. The effects of the inclusion of the effective range are various: the equation of state of a Bose gas undergoes substantial modifications \cite{sala2017, sala2017b, sala2017c, beane2010, beane2018}, and so does the description of the dynamics.
In particular, by considering the effective range contribution in the mean-field dynamics, one obtains the so-called modified Gross-Pitaevskii equation \cite{malomed, gao}, that have been used to predict dynamical signatures of effective range in the case of solitons and sound waves \cite{sgarlata}. Some diffusion Monte Carlo calculations were carried out for studying the validity of a universal description of the bosonic gas, i.e. using the gas parameter $n a_s^3$, where $n$ is the 3D density and $a_s$ the s-wave scattering length. Although in Ref. \cite{giorgini} the universal approach {\cblue is shown to} be valid for usual {\cblue experimental} settings, more recent Monte Carlo investigations with a {\cblue Bose-Bose mixture} \cite{boronat_ultradilute} suggest that {\cblue by increasing the number density} the effective range {\cblue $r_s$ is needed to accurately reproduce the numerical results (see also the analytical results of Ref. \cite{flambaum})}.  

Taking into account that the interaction potential is not {\cblue directly measurable in usual experiments}, in recent years separable potentials \cite{cohen1998} were assumed to investigate nonuniversal features of bosonic and fermionic systems \cite{hammer,braaten,beane2022}. In these papers, which adopt the \gls{eft} methodology, \gls{dr} and \gls{ms} were employed to regularize the divergent loop integrals. In the low-energy limit, which corresponds to consider only s-wave contribution to the phase shift, these calculations are based on the writing of an effective action of which only the terms contributing to the desired low-momentum expansion are retained. This EFT procedure was previously used to study the nucleon-nucleon scattering problem \cite{vanklock, beane_symmetries_2022, kaplan_two-nucleon_1998, birse1999, kaplan1996}. 
It is important to stress that, in the three dimensional case, the nonuniversal EFT corrections of Refs. \cite{hammer,braaten} do not agree with the ones of Refs. \cite{gao,massignan,ketterle, roth2001effective}, 
which are based on the simple Born (zero-order) approximation of scattering theory. {\cblue This disagreement is due to different methods and assumptions in the two approaches. In the first setting \cite{hammer,braaten}, the aim is to obtain the correct low-momentum expansion of the phase shift, by starting from an effective Lagrangian, and summing the Feynman graphs for the $T$-matrix up to the desired momentum power. The second approach \cite{gao,massignan,ketterle,roth2001effective} is instead based on the calculation of the energy shift due to a phase shift in the wavefunction in a finite volume, and then letting the volume go to infinity. A puzzling 
consequence of this energy-shift approach in three spatial dimensions is the fact that sending the s-wave effective range $r_s$ to zero the finite-range correction of the low-momentum expansion {\cred of the interaction potential} remains finite. In this paper we adopt the EFT approach \cite{hammer,braaten} because it is strictly related to the scattering theory via the $T$-matrix, it can be directly applied also to reduced spatial dimensions, and the obtained finite-range corrections are always vanishing for $r_s\to 0$.}

Exact analytical calculations able to tackle the scattering theory by using a realistic finite-range  interaction potential are not available \cite{newton,noyes,kowalski}. In this paper we face this problem 
under the assumption of low energy scattering and using an arbitrary dimension partial wave expansion. Our method, which is based on two crucial approximations on the $T$-matrix equation, called \textit{s-wave} and \textit{on-shell} approximations, allows one to link in a systematic way the s-wave components of the interaction potential and the transition matrix in any spatial dimension $D$. In particular, 
for $D=3$, $D=2$, and $D=1$ we obtain explicit expressions of the low-momentum parameters of the Fourier transform ${\tilde V}(k)$ of the interaction potential $V(r)$ in terms of the s-wave scattering length 
$a_s$ and effective range $r_s$. In this way we recover the nonuniversal EFT results in three spatial dimensions \cite{hammer,braaten}. In the last section, we apply our theory to derive the zero-temperature pressure of the interacting gas in two dimensions in terms of $a_s$ and $r_s$ (see also Ref. \cite{beane2018}).

\section{The two-body problem}

Let us consider the Hamiltonian operator 
\beq 
{\hat H}={\hat H}_0+{\hat V} \; , 
\eeq
where ${\hat H}_0={\hat {\bf p}}^2/(2m_r)$ is the kinetic energy operator 
of a particle of reduced mass $m_r$ and linear momentum ${\hat {\bf p}}$ 
while ${\hat V}$ the interaction potential operator. 
We assume that the potential operator ${\hat V}$ is diagonal 
in the coordinate representation, namely 
${\hat V} |{\bf r}\rangle = V({\bf r}) |{\bf r}\rangle$,
where $|{\bf r}\rangle$ is the eigenstate of the position operator 
${\hat {\bf r}}$, i.e. ${\hat {\bf r}}|{\bf r}\rangle 
={\bf r} |{\bf r}\rangle$. Moreover, $m_r=m/2$ is the reduced mass 
of two identical particles, each of mass $m$. 

\begin{figure}
    \centering
    \includegraphics[width=0.45\textwidth]{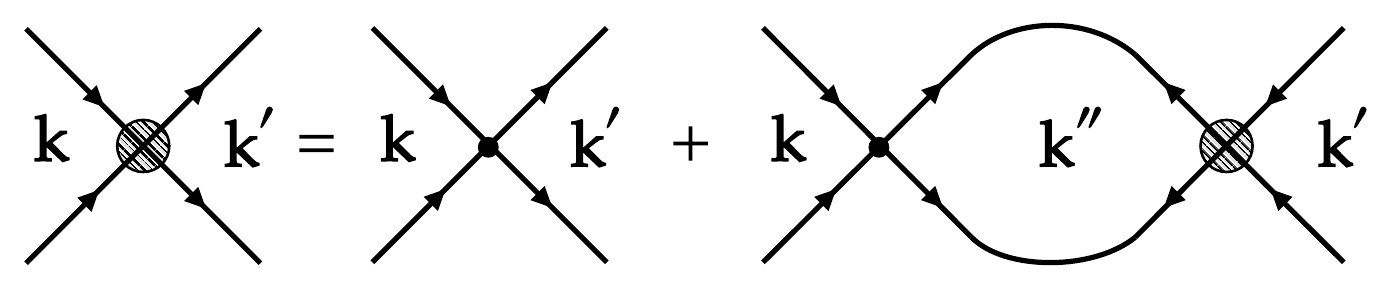}
    \caption{{\cblue Depiction of the scattering process in the center-of-mass reference frame used in the construction of Eq. (\ref{t-equation}). The initial state, represented in the left of all diagrams, is a relative motion in the state $\ket{\mathbf{k}}$, and the final state, in the right side of the diagrams, is in the state $\ket{\mathbf{k'}}$. The shadowed center represents the $T$-matrix, whereas the black dot represents the potential. The last term specifies the relation between initial, final, and intermediate state $\ket{\mathbf{k''}}$.}}
    \label{fig:scattering}
\end{figure}

As shown in many text books \cite{sakurai,stoof,rodberg}, in $D$ spatial dimensions, the matrix element $T_{{\bf k}{\bf k}'}=\langle {\bf k}|{\hat T}|{\bf k}'\rangle$ of the transition operator ${\hat T}$ of scattering theory satisfies the $T$-matrix equation 
\beq 
T_{{\bf k}{\bf k}'} 
= V_{{\bf k}{\bf k}'} + \int d^D{\bf k}'' {V_{{\bf k}{\bf k}''} \over 
{\hbar^2k^2\over 2m_r} - {\hbar^2(k'')^2\over 2m_r} + {\cblue i \, \epsilon} } 
T_{{\bf k}''{\bf k}'} \; , 
\label{t-equation}
\eeq
where $V_{{\bf k}{\bf k}'}=\langle {\bf k}|{\hat V}|{\bf k}'\rangle$, 
$|{\bf k}\rangle$ is the initial state, $|{\bf k}'\rangle$ is the final state, and $|{\bf k}''\rangle$ is an intermediate state. {\cblue The involved variables are represented pictorially in Fig. \ref{fig:scattering}.}

Here, $|{\bf k}\rangle$, $|{\bf k}'\rangle$, and $|{\bf k}''\rangle$ are eigenstates of the linear momentum operator ${\hat {\bf p}}$, i.e. ${\hat {\bf p}}|{\bf k}\rangle = 
\hbar {\bf k} |{\bf k}\rangle$, 
${\hat {\bf p}}|{\bf k}'\rangle = 
\hbar {\bf k}' |{\bf k}'\rangle$, and ${\hat {\bf p}}|{\bf k}''\rangle = 
\hbar {\bf k}'' |{\bf k}''\rangle$. {\cblue In Eq. (\ref{t-equation}), $i$ is the imaginary unit and 
$\epsilon>0$ is an infinitesimal real parameter ensuring that in the scattering there are only outgoing waves.} Notice that 
$V_{{\bf k}{\bf k}'}={\tilde V}({\bf k}-{\bf k}')/(2\pi)^D$, where 
\beq 
{\tilde V}({\bf k}) = \int d^D{\bf r} \ V(r) \ e^{-i{\bf k}\cdot {\bf r}} \; . 
\label{fourier}
\eeq 
is the Fourier transform of the interaction potential $V({\bf r})$. 
We assume that the interaction potential $V({\bf r})$ is spherically symmetric, 
i.e. $V({\bf r})=V(r)$ with $r=|{\bf r}|$, and it follows that ${\tilde V}({\bf k})={\tilde V}(k)$. 

\subsection{Partial waves decomposition}

The $T$-matrix equation (\ref{t-equation}) can be decomposed in partial waves in $D$ dimensions in the following way (see Appendix A for further details). We drop $D$ when the notation is not ambiguous. Define the partial wave expansion of $V_{\mathbf{k}\mathbf{k}'}$ as
\begin{equation}
\label{pwd}
   V_{\mathbf{k}\mathbf{k}'} = \dfrac{1}{(2\pi)^D}\sum_{l}V_l(k, k') \, N(D, l) P_l(\mathbf{\hat{k}}\cdot\mathbf{\hat{k}}'),
\end{equation}
holding also for $T_{\mathbf{k}\mathbf{k}'}$ in an analogous way.
{\cblue The number of spherical harmonics in $D$ dimensions with $D>1$, is the number of independent homogeneous and harmonic polynomials of degree $l$ in $D$ variables, that is \cite{hochstadt}:
\begin{equation}
    N(D, l) = \dfrac{2l+D}{l} \binom{D+l-3}{l-1}.
\end{equation}  
For $D=3$ we recover the usual multiplicities of the spherical harmonics.
It is easy to verify that {\cred when} $l=0$, one has $N(3, 0)=1$, $N(2, 0)=2$.}

Substituting into Eq. (\ref{t-equation}), using the orthogonality of Legendre functions and the uniqueness of the representation in partial waves, one obtains
\begin{equation}
\begin{aligned}
& T_l(k, k') P_l(\mathbf{\hat{k}}\cdot\mathbf{\hat{k}}') = 
V_l(k, k') P_l(\mathbf{\hat{k}}\cdot\mathbf{\hat{k}}') \\
& + \int {d^D\mathbf{k}''\over (2\pi)^D} \Biggl[ \dfrac{N(D,l) }{{\hbar^2k^2\over m} - {\hbar^2(k'')^2\over m}+ {\cblue i \, \epsilon}} V_l(k, k'') T_l(k'', k')\\ & \times P_l(\mathbf{\hat{k}}\cdot\mathbf{\hat{k}}'')P_l(\mathbf{\hat{k}}''\cdot\mathbf{\hat{k}}') \Biggr] \; . 
\end{aligned}
\end{equation}
Choosing $k=k'$, the angular integral can be computed using {\cblue the 
orthonormalization of Legendre polynomials (see Eq. (\ref{orthogonality}) of Appendix B)}.  
By selecting the s-wave term $l=0$ {\cblue we get} 
\begin{equation}
\begin{aligned}
&T_0(k)  =  V_0(k) \\ &+ S_D\int_0^\infty 
{dk''\over (2\pi)^D} \dfrac{(k'')^{D-1}
}{{\hbar^2k^2\over m} - {\hbar^2(k'')^2\over m}
+ {\cblue i\, \epsilon}} V_0(k, k'') 
T_0(k'', k) \; , 
\label{tuttolui}
\end{aligned}
\end{equation}
where $T_0(k)=T_0(k,k)$, 
$V_0(k)=V_0(k,k)$, {\cblue and $S_D = 2 \pi^{D/2}/\Gamma(D/2)$ is the solid angle in
$D$ dimensions with $\Gamma(x)$ the Euler gamma function.}

\section{On-shell approximation}
\label{crux}

Here we adopt the s-wave approximation but also 
the ``on-shell approximation" \cite{rodberg}. 
Explicitly we assume that, due to the singularity in the {\cblue integrand} for $k=k''$, in Eq. (\ref{tuttolui}) $V_{0}(k,k'')\simeq V_0(k,k)=V_0(k)$ and $T_0(k'',k)\simeq T_0(k,k)=T_0(k)$. As a consequence, Eq. (\ref{tuttolui}) becomes
\beq 
T_0(k) = V_0(k) + V_0(k)  \ C(k) \ T_0(k) 
\label{partenza}
\eeq
with 
\beq 
C(k) = S_D\int_0^{\infty} {d{k''}\over (2\pi)^D} 
{1\over {\hbar^2k^2\over m} - {\hbar^2(k'')^2\over m} + {\cblue i \, \epsilon}} \; . 
\label{blubo}
\eeq
Then one finds 
\beq
T_0(k) = {1\over {1\over V_0(k)} - C(k) } \; , 
\label{result}
\eeq
which is the crucial formula of our paper. Eq. (\ref{result}) can be obtained from Eq. (\ref{partenza}) in two ways: by a direct algebraic manipulation or by summing up the associated Born-like geometric series within an iterative scheme. We stress that Eq. (\ref{result}), 
based on the s-wave and the on-shell approximation, is expected 
to be reliable in the regime of low momentum and it becomes 
exact for $k\to 0$ \cite{stoof}. 
Indeed, it turns out that Eq. (\ref{result}) is structurally similar to Eq. (3.16) of Ref. \cite{beane_symmetries_2022}, obtained within the \gls{eft} procedure.

It is important to observe that the s-wave component $V_0(k)$ does not coincide with the Fourier transform ${\tilde V}(k)$. Actually for $D=3$ and $D=2$ we find (see Appendix B) 
\beq 
V_0(k) = \frac{1}{2} \int_0^{\pi} d\theta \ 
{\tilde V}(2k \sin(\theta/2)) \ \sin{(\theta)} \; . 
\label{thisisnew}
\eeq
{\cblue Taylor expanding with respect to $k$ the s-wave component $V_0(k)$ we 
formally obtain 
\beq 
V_0(k) = g_0 + g_2 \ k^2 + ... \; , 
\label{tay}
\eeq
where the coefficients $g_0$ and $g_2$ depend on the choice of the Fourier 
transform ${\tilde V}(k)$ of the interaction potential. Performing also 
the Taylor expansion of the latter, i.e. 
\beq 
{\tilde V}(k) = {\tilde g}_0 + {\tilde g}_2 \ k^2 + ... \; , 
\label{taytilde}
\eeq
one finds, using Eq. (\ref{thisisnew}) that} $g_0 = {\tilde g}_0$ and $g_2 = 2 {\tilde g}_2$. As shown in Appendix B, these two simple relationships are valid also for $D=1$.

\subsection{Dimensional regularization} 

{\cblue In the limit $\epsilon \to 0$}, the term $C(k)$ of Eq. (\ref{blubo}) can be written as 
\beqa 
C(k) &=& - {S_D\over (2\pi)^D} 
{m\over \hbar^2} \int_0^{\infty} dk'' (k'')^{D-1} 
{1\over (k'')^2 + (-i k)^2} 
\nonumber 
\\
&=& - {m\over \hbar^2} (-ik)^{D-2} {B(D/2,1-D/2)\over (4\pi)^{D/2}\Gamma(D/2)} \; , 
\label{betone}
\eeqa 
where $B(x,y)$ is the Euler beta function. Clearly, $C(k)$ 
is ultraviolet divergent at any integer dimension $D$. 
We now show how this divergence is eliminated by \gls{dr} \cite{schakel,sala2016}. 

The Euler beta function 
\beq 
B(x,y) = \int_0^{+\infty} dt {t^{x-1}\over (1+t)^{x+y}}
\eeq
is defined with the real parts of $x$ and $y$ greater than zero. 
However, it can be analytically continued \cite{sala2016} to complex 
values of $x$ and $y$ as
\beq 
B(x,y) = {\Gamma(x) \Gamma(y)\over \Gamma(x+y)} \; . 
\eeq
Performing this analytic continuation in Eq. (\ref{betone}) {\cblue means that we promote the integer spatial dimension $D$ to 
a complex number \cite{schakel,sala2016}. 
After doing it, we can safely {\cred go} back to 
an integer $D$, if $D=3$ and $D=1$ \cite{sala2016}}. {\cblue Thus,} we get \cite{schakel}
\beq 
C(k) = - {m\over \hbar^2} (-ik)^{D-2} 
{\Gamma(1-D/2)\over (4\pi)^{D/2}}  \; ,   
\label{betone-d}
\eeq
where $D$ is {\cblue in general}, {\cblue for the specific discussion of this subsection}, a complex number very close to its integer counterpart. 

From Eq. (\ref{betone-d}), simply setting 
$D=3$ and remembering that $\Gamma(-1/2)=-2\sqrt{\pi}$, we obtain 
\beq 
C(k) = - i k {m\over 4\pi \hbar^2} \; . 
\eeq
Setting $D=1$, and remembering that $\Gamma(1/2)=\sqrt{\pi}$, we have instead 
\beq 
C(k) = - i {1\over k} {m\over 2\hbar^2} \; . 
\eeq

\gls{dr} is more difficult in two spatial dimensions. 
In fact, for $D = 2$ Eq. (\ref{betone-d}) diverges due to the 
presence of $\Gamma(0)$. 
To face this divergence, we extend the calculation to non-integer 
dimension $D = 2-\epsilon$ and let $\epsilon$ go to zero only 
at the end of the calculation. Eq. (\ref{betone-d}) can be written as
\beq 
C(k) = - {m\over \hbar^2} \kappa_0^{\epsilon} (-ik)^{-\epsilon} 
{\Gamma(\epsilon/2)\over (4\pi)^{1-\epsilon/2}}  \; ,   
\eeq
where the regulator $\kappa_0$ is a scale wavenumber which enters for 
dimensional reasons. The small-$\epsilon$ expansion
of the gamma function reads 
\beq 
\Gamma(\epsilon/2) = {2\over \epsilon} - \gamma + O(\epsilon) \; , 
\eeq
where $\gamma\simeq 0.5572$ is the Euler-Mascheroni constant. 
Taking into account that $x^{\epsilon}=e^{\ln(x^{\epsilon})}=e^{\epsilon \ln(x)}=
1+ \ln(x) \epsilon + O(\epsilon^2)$ and $\ln(-i)=-i\pi/2$, we finally get 
\beq 
C(k) = {m\over 2\pi \hbar^2} 
\ln({k\over 2} {e^{\gamma/2}\over \Lambda}) - {m\over 4\hbar^2} i \; ,  
\eeq
after removing the remaining singularity (\gls{ms} scheme) \cite{zeidler} and setting $\Lambda=\sqrt{\pi}\kappa_0$, which plays the role of a ultraviolet cutoff.

\section{Interaction potential and phase shift for $D=3$} 

By using the $D=3$ results of the previous section we can write 
\beq 
T_0(k) = {1\over {1\over V_0(k)} + i k {m\over 4\pi \hbar^2} } \; . 
\label{mystic3D}
\eeq
It is important to underline, that Eq. (\ref{mystic3D}) is a generalization 
of the result obtained in Ref. \cite{braaten} {\cblue with} 
the simple potential $V_0(k)=g_0 + g_2 k^2$. 

A well known result of the scattering theory is that the s-wave 
transition element $T_0(k)$ can be written in term of 
{\cblue the} s-wave scattering amplitude $f_0(k)$ as follows \cite{stoof}
\beq 
T_0(k) = - {4\pi\hbar^2\over m} f_0(k) \; . 
\eeq
Moreover, the s-wave scattering amplitude $f_0(k)$ is related to the s-wave 
phase shift $\delta_0(k)$ by the formula \cite{stoof}
\beq 
f_0(k) = {1\over k \cot(\delta_0(k)) - i k} \; . 
\eeq
Using these two equations with Eq. (\ref{mystic3D}), valid in $D=3$, we get 
\beq 
V_0(k) = - {4\pi \hbar^2\over m} {\tan(\delta_0(k))\over k} \; . 
\label{magic3D}
\eeq
This is our main 3D result: an explicit relationship between $V_0(k)$ of the 3D spherically-symmetric 
interaction potential and the 3D s-wave phase shift $\delta_0(k)$. 
Quite remarkably, Eq. (\ref{magic3D}) is quite similar to the ansatz 
${\tilde V}(k) = - (4\pi \hbar^2/ m) \delta_0(k)/k$ suggested 
in Ref. \cite{massignan}. 

By definition, the 3D s-wave scattering length $a_s$ and the 3D 
s-wave effective range $r_s$ are the low-momenta coefficients of the following 
expansion of the 3D phase shift $\delta_0(k)$ \cite{sakurai,stoof,rodberg}: 
\beq \label{expansion}
k \ \cot({\delta_0(k)}) = - {1\over a_s} + {1\over 2} r_s k^2 + ... \; . 
\eeq
{\cblue This effective range expansion is valid for interaction potentials that decay more rapidly than $r^{-5}$ \cite{massignan}.} Taking into account this low momentum expansion, from  Eq. (\ref{magic3D}) and the Taylor expansion of $V_0(k)$ {\cblue with respect to k}, Eq. (\ref{tay}), we get 
\beq 
g_0 = {4\pi \hbar^2\over m} a_s 
\label{g03D}
\eeq
and 
\beq 
g_2 = {2\pi \hbar^2\over m} a_s^2 r_s \; . 
\label{g23D}
\eeq
Eq. (\ref{g03D}), which relates $g_0$ to $a_s$, is quite familiar 
\cite{sakurai,stoof,rodberg}. Instead Eq. (\ref{g23D}), 
which relates $g_2$ to $a_s$ and $r_s$ is less known, but it can be found in Refs. \cite{braaten,sala2017}. {\cblue Notice that these results, and in particular Eq. (\ref{magic3D}), hold in the regime where $a_s$ is finite while $k$ is small. In other words, Eq. (\ref{magic3D}) cannot be used to model the unitarity regime, where the scattering length $a_s$ diverges, while Eq. (\ref{expansion}) for $a_s=\infty$ and $r_s=0$ simply gives $\delta_0(k)=\pi/2$.} 

\section{Interaction potential and phase shift for $D=1$} 

By using the $D=1$ results of Section \ref{crux} we have 
\beq 
T_0(k) = {1\over {1\over V_0(k)} + i {1\over k}{m\over 2\hbar^2} } \; . 
\label{mystic1D}
\eeq
An interesting achievement of the 1D scattering theory 
is that the s-wave transition element $T_0(k)$ is related to the s-wave 
phase shift $\delta_0(k)$ by the formula \cite{adhikari2000,adhikari2001}
\beq 
T_0(k) = - \left({2\hbar^2\over m}\right) 
\left( {k\over \cot(\delta_0(k)) - i} \right) \; . 
\eeq
Comparing this equation with Eq. (\ref{mystic1D}) we get 
\beq 
V_0(k) = - {2\pi \hbar^2\over m} k \ \tan(\delta_0(k)) \; . 
\label{magic1D}
\eeq
This is our main 1D result: an explicit relationship between the 
Fourier transform $V_0(k)$ of the 1D spherically-symmetric interaction potential 
and the 1D s-wave phase shift $\delta_0(k)$. 

By definition, the 1D s-wave scattering length $a_s$ 
and the 1D s-wave effective 
range $r_s$ are the low-momenta coefficients of the following 
expansion of the 1D phase shift $\delta_0(k)$ \cite{adhikari2000,adhikari2001} 
\beq 
k \ \tan({\delta_0(k)}) = {1\over a_s} + {1\over 2} r_s k^2 + ... \; . 
\eeq
Taking into account this low momentum expansion, from Eq. (\ref{magic1D}) and the Taylor expansion 
of $V_0(k)$, Eq. (\ref{tay}), we obtain 
\beq 
g_0 = - {2\hbar^2\over m a_s} 
\label{g01D}
\eeq
and 
\beq 
g_2 = - {\hbar^2\over m} r_s \; . 
\label{g21D}
\eeq
Eq. (\ref{g01D}), which relates $g_0$ to $a_s$, is quite familiar 
\cite{adhikari2000,adhikari2001}. Instead Eq. (\ref{g21D}), 
which relates $g_2$ to $r_s$, was previously found in Ref. \cite{sala2017b}. 

\section{Interaction potential and phase shift for $D=2$} 

By using the $D=2$ results of Section \ref{crux} the s-wave transition 
element reads 
\beq 
T_0(k) = {1\over {1\over V_0(k)} - {m\over 2\pi \hbar^2} 
\ln({e^{\gamma/2} k\over 2\Lambda}) + {m\over 4\hbar^2} i } \; . 
\label{mystic2D}
\eeq
In the 2D scattering theory the s-wave transition element $T_0(k)$ 
is related to the s-wave phase shift $\delta_0(k)$ by the 
formula \cite{adhikari1986}
\beq 
T_0(k) = - \left( {4\hbar^2\over m}\right) 
\left( {1\over \cot(\delta_0(k)) - i} \right) \; . 
\eeq
Comparing this equation with Eq. (\ref{mystic2D}) we find 
\beq 
V_0(k) = - \left( {4 \hbar^2\over m}\right) {1\over \cot(\delta_0(k)) 
- {2\over \pi} \ln( {k\over 2} {e^{\gamma/2}\over \Lambda})}  \; . 
\label{magic2D}
\eeq
This is our main 2D result: $V_0(k)$ of the 2D spherically-symmetric interaction potential 
in terms of {\cblue the} 2D s-wave phase shift $\delta_0(k)$. 
Eq. (\ref{magic2D}) clearly depends on the ultraviolet cutoff $\Lambda$. 

By definition, for short-range potentials, the 2D s-wave scattering length $a_s$ 
and the 2D s-wave effective range $r_s$ are the coefficients of the following 
low momentum expansion of the 2D phase shift $\delta_0(k)$ \cite{khuri} 
\beq 
\cot(\delta_0(k)) = {2\over \pi} \ln({k\over 2} a_s e^{\gamma}) 
+ {1\over 2} r_s^2 k^2 + ... \; . 
\label{bubo}
\eeq
Inserting this expression into Eq. (\ref{magic2D}) 
we obtain 
\beq 
V_0(k) = -\left( {4\hbar^2\over m}\right) {1\over {2\over \pi} \ln(\Lambda a_s e^{\gamma/2})+{1\over 2} r_s^2 k^2+...} \; ,  
\label{rong}
\eeq
which, remarkably, does not have anymore a logarithmic dependence on $k$ and it 
is convergent for $k\to 0$. We can then write the low momentum expansion of $V_0(k)$, given by Eq. (\ref{tay}), finding 
\beq 
g_0 =-{4\pi\hbar^2\over m} {1\over \ln( \Lambda^2 a_s^2 e^{\gamma})} \; . 
\label{g02D}
\eeq
This result is consistent with the one obtained by Castin \cite{castin-obscure}. We also obtain the formula 
\beq 
g_2 = {2\pi^2 \hbar^2\over m} {r_s^2\over 
\ln^2(\Lambda^2 a_s^2 e^{\gamma})} 
\label{g22D}
\eeq
which relates $g_2$ to the s-wave scattering 
length $a_s$, the effective range $r_e$ and the cutoff $\Lambda$. Sometimes in many-body calculations it is used some other characteristic range $R$ of the inter-atomic potential $V(r)$ instead of the effective range $r_s$ \cite{boronat,sala2017c,sala2018}. 

{\cblue For ease of reading, we summarize the results for all the dimensions in Tab. (\ref{tab:summary}), reporting the function $C(k)$, and the low-momentum coefficients $g_0$, $g_2$.}
\begin{table}[]
\cblue{
    \renewcommand{\arraystretch}{2}
    \centering
    \begin{tabular}{|c|c|c|c|}
    \hline
     $D$ & $C(k)$ & $g_0$ & $g_2$ \\
    \hline
    \hline
     $3$ & $- i k {m\over 4\pi \hbar^2} $& ${4\pi \hbar^2\over m} a_s$ & ${2\pi \hbar^2\over m} a_s^2 r_s$ \\
     $2$ & ${m\over 2\pi \hbar^2} \ln({k\over 2} {e^{\gamma/2}\over \Lambda}) - {m\over 4\hbar^2} i$ & $-{4\pi\hbar^2\over m} {1\over \ln( \Lambda^2 a_s^2 e^{\gamma})}$ & ${2\pi^2 \hbar^2\over m} {r_s^2\over 
\ln^2(\Lambda^2 a_s^2 e^{\gamma})} $\\
     $1$ & $ -i {1\over k} {m\over 2\hbar^2}$  & $- {2\hbar^2\over m a_s}$  & $- {\hbar^2\over m} r_s$\\
     \hline
\end{tabular}
    \caption{Main results in spatial dimension $D=1,2,3$. 
    $C(k)$ is the function of the $T$-matrix derived using dimensional regularization, $g_0$ and $g_2$ are the first two 
    coefficient of the low-momentum expansion of the s--wave component $V_0(k)$ of the interaction potential. The tabulated quantities depend on the s-wave scattering length $a_s$ and the s-wave effective range $r_s$. For $D=2$ there is also a dependence on the wavenumber ultraviolet cutoff $\Lambda$.}
    }
    \label{tab:summary}
\end{table}

\section{An application: Effective field theory of interacting bosons}

The formalism developed in the previous Sections is well suited to setup low-momenta \gls{eft}s of bosons and fermions. 
As an example, let us consider the Lagrangian density of identical bosonic particles of mass $m$ in a $D$ spatial dimensions, given by 
\beqa
\mathcal{L} &=& \psi^*({\bf r},t) \bigg[ i \hbar\partial_{t} + \frac{\hbar^2}{2m} \nabla_{\bf r}^2 \bigg]\psi({\bf r},t) 
\nonumber 
\\
&-&\frac{1}{2}\int d^D{\bf r}' |\psi({\bf r}',t)|^2 V(|{\bf r}-{\bf r}'|)|\psi({\bf r},t)|^2 \; ,
\label{lagrangian}
\eeqa
where the bosons are described by the complex field $\psi(x,\tau)$ and $V(|x-x'|)$ is the two-body interaction 
potential between atoms. By using Eq. (\ref{taytilde}) it is straightforward (see for instance \cite{sala2017,sala2017b,sala2018}) to get the low-momenta effective Lagrangian density 
\beqa
\mathcal{L} &=& \psi^*({\bf r},t) \bigg[ i \hbar\partial_{t} + \frac{\hbar^2}{2m} \nabla_{\bf r}^2 \bigg]\psi({\bf r},t) 
-\frac{1}{2} {\tilde g}_0 |\psi({\bf r},t)|^4 
\nonumber 
\\
&+& {1\over 2} {\tilde g}_2 |\psi({\bf r},t)|^2 \nabla_{\bf r}^2 |\psi({\bf r},t)|^2 \; . 
\label{lagrangian-local}
\eeqa
Quite remarkably, contrary to Eq. (\ref{lagrangian}), the effective Lagrangian density of Eq. (\ref{lagrangian-local}) is local. 
The connection with the scattering theory is established by the formulas of ${\tilde g}_0$ and ${\tilde g}_2$ as a function of the s-wave scattering length $a_s$ and the s-wave effective range $r_s$. As previously stressed, while ${\tilde g}_0$ coincides with $g_0$, ${\tilde g}_2$ differs from $g_2$ by a factor 2 in any dimensions $D$. Instead, the connecting formulas are crucially dependent on $D$. 
{\cblue Formally, the modified Gross-Pitaevskii equation,  
derived as Euler-Lagrange equation from the effective Lagrangian density (\ref{lagrangian-local}), is equivalent to  the one found by several authors \cite{gao, malomed, massignan},} {\cred but it contains a coefficient $\tilde{g}_2$ which is related in a different way to scattering parameters, as discussed previously.}

\begin{figure}
    \centering
    \hspace{-10pt}
    \includegraphics[width=0.48\textwidth]{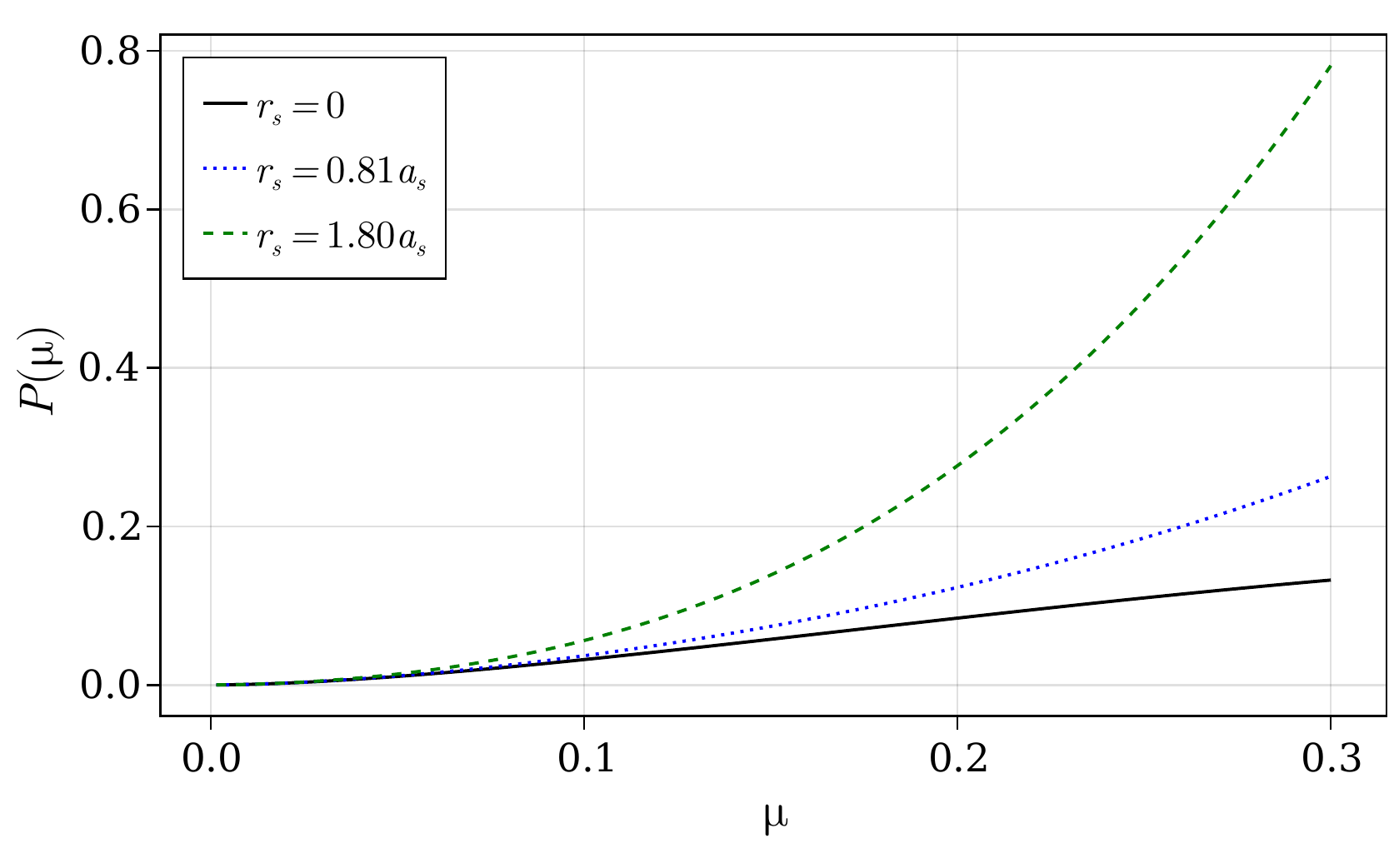}
    \caption{{\cblue
Zero-temperature pressure dependence on chemical potential, for three values of effective range $r_s$. Values are as computed in Ref. \cite{flambaum}. Dashed line is for case of Li-Li singlet state scattering, dotted line is for the case of Na-Na triplet state scattering, solid line is the prediction with the zero-range model. Length is in units of $a_s$, energy is in units of $\hbar^2/(ma_s^2)$.}}
    \label{fig:pressure}
\end{figure}

As we have seen, the case $D=2$ is quite complicated  because ${\tilde g}_0$ and ${\tilde g}_2$ depend on the ultraviolet cutoff $\Lambda$. We now show that, quite remarkably,  this cutoff can be washed out in explicit  calculations. For instance, at one-loop level, from Eq. (\ref{lagrangian-local}) one finds \cite{sala2017c,sala2018,tononi2019}, after  \gls{dr}, the following expression for the zero-temperature pressure $P$ of the interacting Bose gas as a function of the chemical potential $\mu$: 
\beqa
P(\mu) &=& {1\over 2 {\tilde g}_0} \mu^2 
+ {m\over 8\pi\hbar^2} {\mu^2\over 
(1+ {4m \, {\tilde g}_2\over \hbar^2 {\tilde g}_0}\mu )^{3/2}} 
\nonumber 
\\
&\times&
\ln{\big({4\hbar^2 \Lambda^2
\over m \mu e^{\gamma+1/2}} 
(1+ {4m \, {\tilde g}_2\over \hbar^2 {\tilde g}_0}\mu )\big)}
\; , 
\label{delirioserale}
\eeqa
where the first term is the mean-field result 
and the second one is the Gaussian (one-loop) correction with $\Lambda$ the same ultraviolet cutoff of Eq. (\ref{g02D}) and $\gamma$ the Euler-Mascheroni constant. Contrary to Refs. \cite{sala2017c,sala2018,tononi2019}, here we  explicitly use both Eqs. (\ref{g02D}) and (\ref{g22D}). Inserting these equations into Eq. (\ref{delirioserale}) we obtain 
\beq
P(\mu) = 
{m\over 8\pi\hbar^2} \mu^2 \ln{\left( {4\hbar^2\over m \mu a_s^2 e^{2\gamma+1/2}}\right)} + {3 m^2\over 16\hbar^4} r_s^2 \ \mu^3 \; ,  
\label{sivedra}
\eeq
where the first term is $\Lambda$ independent while the 
second term is obtained in the limit $\Lambda \to +\infty$. For $r_s=0$ our result for the pressure $P(\mu)$, derived with \gls{dr}, becomes  the same of that found by Mora and Castin \cite{mora2003,mora2009} with space discretization. For $r_s\neq 0$ Eq. (\ref{sivedra}) is fully consistent with the EFT findings of Ref. \cite{beane2018}. {\cblue The zero-temperature pressure is represented in Fig. \ref{fig:pressure} for three values of $r_s$, corresponding to the case $r_s=0$, and two values computed in Ref. \cite{flambaum} using the van der Waals model for Li-Li and Na-Na scattering.}

\section{Conclusions}

We have shown a method for systematically relate, in generic spatial dimension $D$, the coefficients of the low momentum expansion of the interaction potential in terms of the s-wave scattering length and effective range, highlighting the two crucial assumptions that are present in the scheme, namely the s-wave and the on-shell approximations for the $T$-matrix equation. The on-shell approximation turns out to be an alternative to the assumption of a separable potential utilized in previous works \cite{cohen1998}.
We have explicitly calculated these relations in dimensions $D=1, \ 2, \ 3 $ by using dimensional regularization, and we also discussed the discrepancy appearing in the literature for the expression of $g_2$ in the $D=3$ case, showing, using a different method, how the same results of Ref. \cite{hammer, braaten} can be obtained.
Using this framework we have also obtained the finite-range correction to the zero-temperature pressure in a $D=2$ Bose system, which is in agreement with previous results \cite{beane2018}.
It may be interesting to extend the proposed scheme for the case of atomic mixtures, which is getting increasing interest, 
or for atomic Josephson junctions in reduced spatial dimensions. 

\section*{Acknowledgements}

The authors thank S.K. Adhikari, G. Bertaina, A. Cappellaro, L. Dell'Anna, and A. Tononi for useful comments and suggestions. This work has been partially supported by the  
Iniziativa Specifica ``Quantum'' of INFN, 
by the BIRD grant ``Ultracold atoms in curved geometries'' of the University of Padova, 
{\cblue and by the European Union-NextGenerationEU within the National Center for HPC, Big Data and Quantum Computing (Project No. CN00000013, CN1 Spoke 1: “Quantum Computing”)}.

\section*{Appendix A: Legendre polynomials in arbitrary dimensions}
Legendre polynomials in arbitrary dimension are defined, after fixing a direction, as the spherical hamonic of a rotational invariant homogeneous hamonic polynomial with respect to this direction. This defines them in a unique way \cite{hochstadt}. We indicate them with the notation $P_l(\cdot)$ when the dimension is obvious.
$P_l(\cdot)$ will in general depend on two versors, but due to the rotational invariance, it is only dependent on the angle in between through the inner product of versors. For every versor  $\hat{\mathbf{x}}$ the following normalization condition holds 
\begin{equation}\label{orthogonality}
    \int_{\mathbb{S}_D} d\omega \, P_l(\hat{\mathbf{x}}(\omega)\cdot\hat{\mathbf{y}})P_{l'}(\hat{\mathbf{x}}(\omega)\cdot\hat{\mathbf{y}}) = \dfrac{S_D}{N(D, l)} \delta_{ll'},
\end{equation}
where $\mathbb{S}_D$ is the unit spherical shell in $D$ dimensions, and $\hat{\mathbf{x}}$ is the corresponding versor.
It is important to notice that it holds
\begin{align}
    P_l(1)=&1, \label{prop1} \\
    P_l(-1)=&(-1)^l. \label{prop2}
\end{align}
The above condition allows one to define the Legendre polynomial in the case $D=1$. In this case the angle can only assume values $0$ or $\pi$. Let $\hat{\mathbf{u}}^+$ be the versor in the positive direction, and $\hat{\mathbf{u}}^-$  in the negative direction. Integrating in the discrete measure, for $l=l'$
\begin{equation}
\int_{\mathbb{S}^1} d\omega \, P_l^2(\hat{\mathbf{x}}(\omega)\cdot\hat{\mathbf{y}}) = P_l^2(\hat{\mathbf{u}}^+\cdot\hat{\mathbf{y}}) + P_l^2(\hat{\mathbf{u}}^-\cdot\hat{\mathbf{y}}) = 2.
\end{equation}
in the last equality we used the fact that the argument of the Legendre polynomials can only assume values $\pm 1$, and the properties (\ref{prop1}) and (\ref{prop2}).
Remembering that $S^1 = 2$, and using Eq. (\ref{orthogonality}) we define the value of $N(1, l) := 1$.
By using the rotational symmetry, integral (\ref{orthogonality}) can be evaluated separately in the angular variables that fixes the inner product. Let $\hat{\mathbf{x}}(\omega)\cdot\hat{\mathbf{y}} = t$,
\begin{align*}
    &\int_{\mathbb{S}_D} d\omega \, P_l(\hat{\mathbf{x}}(\omega)\cdot\hat{\mathbf{y}})P_{l'}(\hat{\mathbf{x}}(\omega)\cdot\hat{\mathbf{y}})= \\ &= S_{D-1}\int_{-1}^1 P_l(t)P_{l'}(t) (1-t^2)^{(D-3)/2},
\end{align*}
obtained by using the spherical hypersurface of radius $\sqrt{1-t^2}$ in $D$ dimensions:  $S_D(1-t^2)^{(D-3)/2}$ \cite{hochstadt}.
This normalization condition will be used in computing the partial wave expansion used in the s-wave approximation.

Finally, we point out that a similar generalization is available also for spherical Bessel functions, which are coefficient of the radial component of the partial wave expansion of the plane wave in general dimension $D$, are defined as \cite{stoof1988,hochstadt}
\begin{equation}
    j_{l, D} (z) = \Gamma\left(\dfrac{D}{2}\right) \left(\dfrac{2}{z}\right)^{\frac{D}{2}-1} J_{l+\frac{D}{2}-1}(z)
\end{equation}
where $J_\alpha$ is the Bessel J function of index $\alpha$, that can be rational.

\section*{Appendix B: Connection between s-wave and Fourier transform}

As explained in the Introduction, the representations in momentum space of the matrix element of the operators $\hat{T}$ and $\hat{V}$ take the form of {\cblue Fourier transforms} calculated in the difference between the wavevectors. Let us focus on the operator $\hat{V}$, since the treatment of the operator $\hat{T}$ is identical. The Fourier transform is denoted by $\tilde{V}(\textbf{k}-\textbf{k}')$, 
\begin{equation}\label{matrix_element}
    V_{\textbf{k}\textbf{k}'} = \dfrac{\tilde{V}(\textbf{k}-\textbf{k}')}{(2\pi)^D}=\dfrac{1}{(2\pi)^D}\int d^D \textbf{r} \ V(r) \  e^{-i(\textbf{k}-\textbf{k}') \cdot \textbf{r}},
\end{equation}
In the hypothesis $|\textbf{k}| = |\textbf{k}'|=k$, the difference vector can be expressed as
\begin{equation}
    \textbf{k}-\textbf{k}' = 2k\sin\left(\theta/2\right) \hat{\textbf{u}},
\end{equation}
where $\hat{\textbf{u}}$ is the versor of the difference, and $\theta$ the angle between the wavevectors. Clearly, for $D=1$ the angle $\theta$ has only two values: $\theta=0$ and $\theta=\pi$. 
It follows that the expression $\tilde{V}(\textbf{k}-\textbf{k}')$ only depends on $k$ and $\theta$, so we refer to this quantity with the notation $\tilde{V}(2k\sin(\theta/2))={\cblue {\tilde V}({\bf k}-{\bf k}')}$. 

By using the standard expansion in partial waves, i.e. the Fourier-Legendre series, 
we can write 
\begin{equation}\label{legendre}
    \tilde{V}(2k\sin(\theta/2)) = \sum_{l=0}^{\infty} V_l(k)(2l+1)P_l(\cos(\theta)),
\end{equation}
where $P_l(x)$ are Legendre polynomials, that satisfy the orthogonality relation \cite{NIST}
\begin{equation}
    \int_{-1}^{1} dx \ P_{l}(x)P_{l'}(x) = \dfrac{2}{2l+1}
    \delta_{l,l'} \; . 
\end{equation}
As a direct consequence of the Fourier-Legendre expansion one can compute the expansion coefficients via integration. {\cblue These integrals are convergent for 
potentials that are square summable (Fischer-Riesz theorem). Explicit examples of potentials that satisfy this condition are discussed, for instance, in Refs. \cite{stoof,boronat2023}.} 
In 3D and 2D, the integration is simply
\beq 
V_l(k) = {1\over 2} 
\int_0^{\pi} d\theta \ 
{\tilde V}(2k\sin(\theta/2)) 
\ \sin(\theta) \ 
P_l(\cos(\theta)) \; . 
\eeq
The s-wave case, i.e. $l=0$, 
gives exactly Eq. (\ref{thisisnew}) because $P_0(x)=1$. 
However, in 1D the set of angles that $\theta$ can assume is discrete, containing only in $0$ and $\pi$. The same integral can be evaluated in a discrete measure giving 
\begin{equation}
    V_l(k) = \dfrac{1}{2} \left(\tilde{V}(0) + (-1)^l\tilde{V}(2k)\right) \; .
\end{equation}
Notice that the first term of the expansion is the even part of $\tilde{V}$ with respect to the variable $\theta$ centered in $\pi/2$.
Independently of the dimension D, the relationships $g_0 = {\tilde g}_0$ and $g_2 = 2 {\tilde g}_2$ are verified.

\end{document}